\documentclass[pdflatex,sn-mathphys]{sn-jnl}

\theoremstyle{thmstyleone}%

\theoremstyle{thmstyletwo}%

\theoremstyle{thmstylethree}%

\usepackage{graphics}
\usepackage{xcolor}
\usepackage{caption}
\begin{document}

\title{Mood Classification of Bangla Songs Based on Lyrics}


\author[1]{\fnm{Maliha} \sur{Mahajebin}}
\author[2]{\fnm{Mohammad Rifat Ahmmad } \sur{Rashid}}
\author[1]{\fnm{Nafees} \sur{Mansoor}}


\affil[1,3]{\orgname{University of Liberal Arts Bangladesh}, \orgaddress{\city{Dhaka}, \country{Bangladesh}}\email{malihamahajebin1@gmail.com}}
\affil[2]{\orgname{East West University, Bangladesh}, \orgaddress{ \city{Dhaka}, \country{Bangladesh}}\email{rifat.rashid@ewubd.edu}}
\affil[]{\email{nafees.mansoor@ulab.edu.bd}}

\maketitle


\abstractname{Music can evoke various emotions, and with the advancement of technology, it has become more accessible to people. Bangla music, which portrays different human emotions, lacks sufficient research. The authors of this article aim to analyze Bangla songs and classify their moods based on the lyrics. To achieve this, this research has compiled a dataset of 4000 Bangla song lyrics, genres and used Natural Language Processing and the Bert Algorithm to analyze the data. Among the 4000 songs, 1513 songs are represented for sad mood, 1362 for romantic mood, 886 for happiness, and the rest 239 are classified as relaxation. By embedding the lyrics of the songs, the authors have classified the songs into four moods: Happy, Sad, Romantic, and Relaxed. This research is crucial as it enables a multi-class classification of songs' moods, making the music more relatable to people's emotions. The article presents the automated result of the four moods accurately derived from the song lyrics.}




\section{Introduction}
Music is the only language that everyone can understand. It plays a very important role for all people around the world. People can’t imagine a party, movie, or event without music. Researchers in \cite{bib1} find that music enhances brain functioning. Due to the regulation of dopamine in the brain, sounds like music and noise have a major impact on our moods and emotions. Dopamine in the brain is a neurotransmitter that actively participates in emotional behavior and mood regulation With the abundant resources of easily-accessible digital music libraries and online streaming of music over the past decade, research on automated systems of music is very noticeable as well as challenging in this era\cite{bib2}. Some common search and retrieval categories are- artist, genre, mood, lyrics are more. Though the music itself is the expression of various emotions, which can be highly subjective and difficult to quantify.\par
The automated classification method is to identify the mood of the songs referring to human emotions\cite{bib3}. When something has a great relationship with one’s mood to be happy, he can listen to happy songs. On the other hand, when one is going through a difficult time or losing someone, it is helpful to listen to sad songs that make the person feel much better. One can listen to relaxation songs in bed which will help to have a good sleep. While listening to sad or happy songs many applications recommend similar mood-type songs. This feature is growing and gaining popularity far and wide. 
Much research has already been done for the classification of English songs by using mood. On the other hand, few works are done for classifying Bangla music based on mood.\par 
Bangla music has a very rich music culture, though there are only a few prominent works done for Bangla music classification based on mood. Research has been done on Bangla audio songs, however, the number of research  based on lyrics is not many. Bangla songs have numerous emotions though the researchers are limited to classifying the songs into happy and sad moods only.\par
Hence, the objective of this research is to detect the various moods of Bangla songs based on lyrics. The aim of this article is to create an application that classifies and predicts songs' moods by analyzing the lyrics. The main motivation for this project is to build software that will help people to listen to songs depending on their mood. In this article, the authors have tried to classify both audio and lyrics based on songs. This research proposes a multi-class classification model by using BERT(Bidirectional Encoder Representation from Transformers )\cite{bib4}, a very popular NLP(Natural Language Processing) sequence-to-sequence model. The authors are using the deep learning library Keras to classify our five different moods as Romantic, sad, happy, religious, and patriotic. Google made the BERT model as an open source to improve the efficiency of NLP and the efficiency is much higher than the other classification model like naive Bayes and logistic regression. This research is using the BERT model, TensorFlow hub, pre-training, small BERTs, Kaggle, and Adam(Adaptive moments) used for fine-tuning.\par
This research work is focused on mood analysis using Bangla songs' lyrics. In This research, a dataset containing 4000 Bangla Songs' lyrics has been prepared and presented. The research has also performed an exploratory data analysis which helps to analyze the data. The BERT algorithm is used for mood classification in this research.

The rest of the paper is organized as follows. Section 2 describes the existing research works regarding Music mood classification. The prepared dataset and data analysis are presented in Section 3. A discussion of the methodology and the algorithm is presented in Section 4, and the experimental results, as well as the conclusion, are presented in Section 5.

\section{Related Work}\label{sec2}
Sentiment analysis is an epoch-making topic in this era. Enormous research has been done and still going on this topic. Music analysis, a significant part of sentiment analysis, has earned the interest as well as the focus of the researchers. Many researchers have done varieties of works in the field of Music analysis. Some research has been done on the basis of the audio signal, some on the basis of Lyrics, and also the classification has been done on different topics. Such as artists, categories, moods of the songs, and so on.\par
A huge survey was done on sentiment analysis’s recently updated algorithms in \cite{bib5} by some students of Ain Shams University in 2014. The researchers analyzed 54 articles and the result was generated by analyzing open-field research that SC and FS algorithms have more enhancements. Machine Learning algorithms, most frequently used to solve SC problems are- Native Bayes and Support Vector Machines. It was also analyzed that most of the papers on the English language, though the research on other languages is growing up.\par
Another survey on the challenges of Sentiment analysis was done by Hussein, et al. in 2016. In \cite{bib6} the work was based on 47 papers discussing the importance and effects of sentiment 191 analysis challenges in sentiment evaluation. The research was based on two com- 192 parisons. The result of this analysis shows another essential factor to recognize the 196 sentiment challenges. This is domain dependence. In all types of reviews, the popular challenge is the negation challenge.\par
Regarding music analysis research work, important research had been done on Bangla song Reviews to research the acceptance of a young star’s song. The authors collected the comments from the comment section of a Youtube music video of a Bangla young star to classify if the comments were positive or negative. A lexicon-based backtracking approach on each sentence of the dataset was used in \cite{bib7}.\par 
Some students of Daffodil International University Bangladesh had also done some research on Bangla Music. The researchers had done Bangla song genre classification using a neural network. In \cite{bib8} a deep learning model was proposed using the Sequential using Sequential Model of the deep learning library Keras in order to classify 6 different Bangla music genres. The researchers used audio signals for that purpose and collected an MP3 dataset of Bangla songs.\par
Automatic mood classification using tf*idf based on music lyrics was the earliest work done by van Zaanen, M.; Kanters, P.H.M. in 2010. The research divided the mood of songs into different classes such as sad, happy, angry, relaxed, etc \cite{bib9}. The authors used tf*idf which helped to emphasize the expressive words of the lyrics of a song and automatically describe the mood.\par
Mood-based Classification of Music analyzing lyrical data through text mining was done by Amity University Uttar Pradesh, Noida – INDIA. The researchers also used \cite{bib10} Naive Bayes classifier as well as some audio features of the song for text mining and getting the mood of the song. Sebastian Raschka had done some work on Predicting the mood of music from song lyrics using the Bayes classifier\cite{bib11}. The researcher developed a web app to perform mood prediction by giving the artist's name and song title. A huge number of songs from the ‘Million song dataset’ was collected and trained using Bayes theorem classifying the songs in two moods- happy and sad. The results had shown that a naive Bayes model can predict the positive class with high precision. To filter a large music library for happy music with a low false positive rate, this model is useful.\par 
Similar work is done on a number of Bangla song lyrics by some students of United International University Bangladesh. In the research, \cite{bib12} the authors worked on Music mood classifiers on Bangla song lyrics using Native Naive Bayes classifier. The work also divided the Bangla songs into two moods- sad and happy. An optimal result showed that the sad songs are increasing day by day as peoples’ emotions are going down with the era.\par

Most of the work reported in the literature regarding music analysis was limited to classifying the songs in only two moods as well as most of the lyrical works are based on the tf*idf algorithm. In this research, the authors have collected extensive datasets and used those data for making a prediction using the Bert algorithm. This work will do a multiclass classification of the songs classifying them into four moods. A dataset of 4000 Bangla songs will definitely help to generate more accurate predictions of different moods and more precise classification of the songs’ moods compared to other reported works.

\section{Proposed Work}\label{sec4}

Alike any other intelligent prediction system, a dataset carries an important role in the Bangla music mood analysis. To define the moods a proper dataset was needed which will contain varieties of Bangla music. As this research needs the dataset very much, a dataset has been created and analyzed in the latter part of this article.

\subsection{Dataset Description}\label{subsec2}
In this machine learning system, the system gains knowledge from the dataset from the lyrics of the songs and predicts the song's mood. The dataset used in this research is a collection of 4000 Bangla songs. In this research Bangla songs' lyrics have been scraped from different Bangla song lyrics websites which include banglasongslyrics.com, Bengali Lyrics(gdn8.com), etc. Apart from the scrapped lyrics, other lyrics have been incorporated into the dataset manually. Since Bangla Music has abounded emotions, this research has collected more than twenty categories of songs from different writers as well as genres in figure 1. In this research, songs have been collected from the last 50 years' time frame. Though 50 years time frame is considered for the research, some old singers hold a large field in the Bangla music industry, such as Rabindranath Tagore, Kazi Nazrul Islam, Lalon shah, and so on. The dataset contains some of their music. In the dataset, there are also tribal songs, modern songs, songs from cinema, and so on. The highest amount of songs is from Rabindranath Tagore as his songs carry lots of emotions and the lyrics are helpful as training data. There are around 25 songs in the dataset that are uncategorized.\par

\begin{figure}[htpb]
\centering
\includegraphics[width=3.5in, height=2.3in]{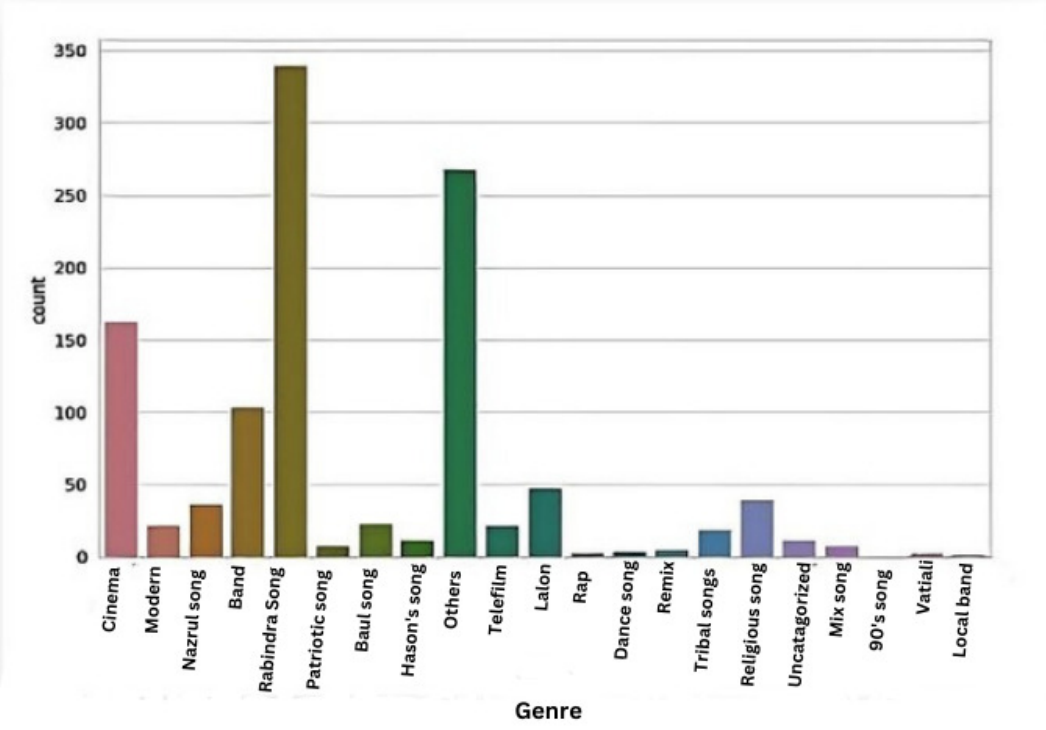}
\caption{Picture of music categories in dataset}\label{fig:1}
\end{figure}
To process, the authors have scaled those data. Moreover, the initial data has been cleaned, because there
were many null values. Besides, for some songs the lyrics cannot be fully collected, it happened in the case of old songs especially. Finally, the authors compiled all the data to apply the machine learning algorithms and composed the compiled Excel format into a
comma-separated value (CSV) format. There are four columns in the dataset- title, category, lyrics, and mood. The title is the song’s name published by the copyrighted author. 
Among the four columns of the dataset, the lyrics column holds great value. In the lyrics column, there is the full version of the song’s lyrics. Those lyrics are processed for the prediction system. The mood column is an essential part of this dataset as in this article moods will be predicted. There are four Bangla songs moods identified from various Bangla and English songs mood identification research done previously\cite{bib13}. The four moods classified are romantic, sad, happy, and relaxed. After collecting all the songs, the dataset according to the four moods- happy, sad, romantic, and relaxed has been labeled. There is some unbalance in the dataset because there are little amount of relaxation songs the authors have found from Bangla music history. On the contrary, a huge number of songs contain both happy as well as romantic emotions.

\subsection{Mood Categories in Dataset}\label{subsec2}
The dataset is trained to predict the mood of the song. Analyzing the entries in the dataset which is more than 16000, the researchers have selected four kinds of moods-romantic, sad, happy, and relaxed. Though there is some unbalance in the dataset due to the confusing emotion of the songs. Fig 2 shows the percentage of how the songs are divided into the four moods. In figure 2 the number of songs in different moods is also visible. The majority of the songs in the dataset are sad. There are 1513 songs of sad mood which holds the largest percentage in the dataset. There are 1362 romantic songs, 886 happy songs, and the rest 239 relaxed songs. All of the 4000 songs have been labelled for training purposes. Romantic and happy songs' lyrics have some words in common, so these two moods have been difficult to label in the dataset. It is difficult to find relaxation (100) song lyrics from Bangla songs documentaries as there are few songs holding this particular mood. The authors are still trying to increase the dataset and balance the dataset to get a good performance.\par
\begin{figure}[htpb]
\centering
\includegraphics[width=3in, height=2.2in]{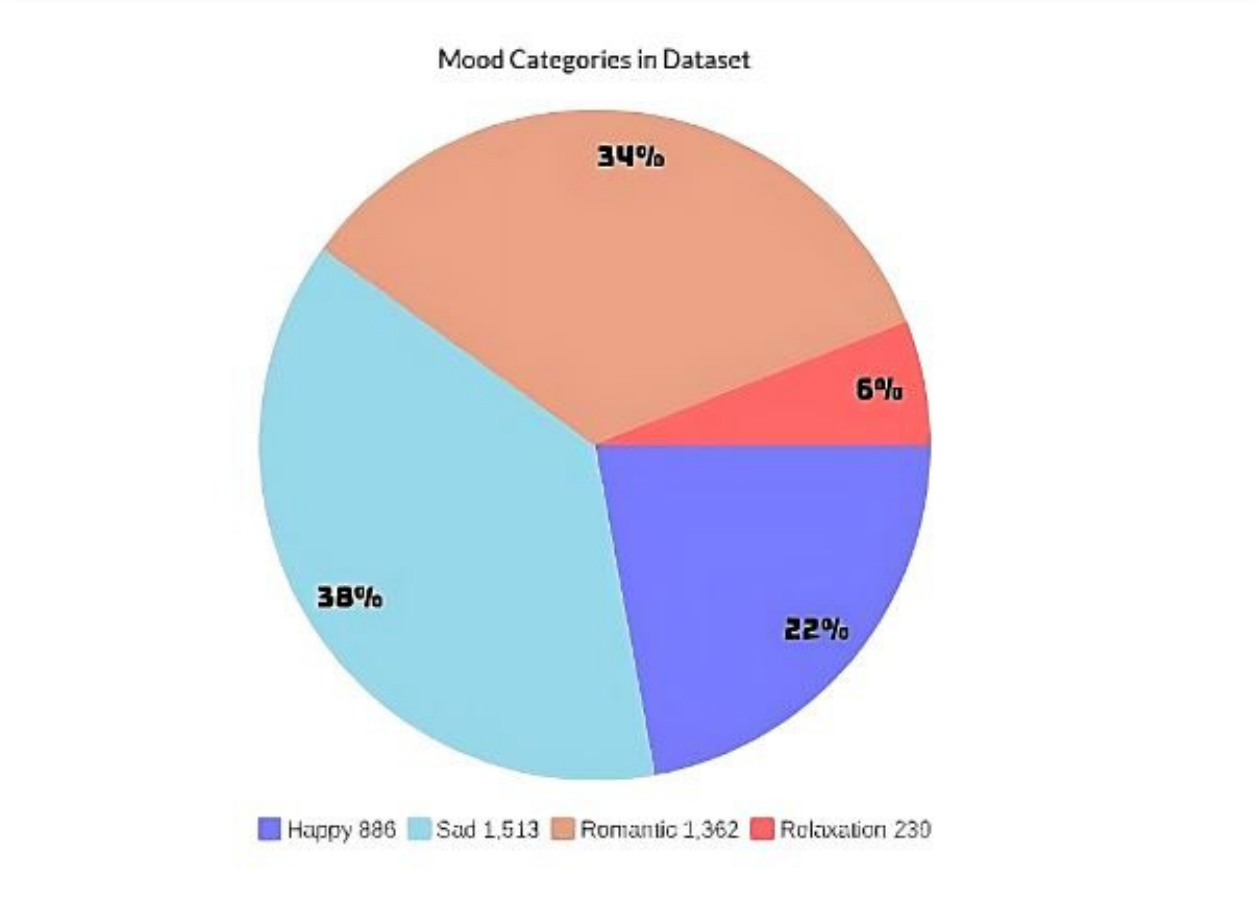}
\caption{Mood categories in dataset}\label{fig:2}
\end{figure}
In order to analyze the dataset, the authors have used word tokenization. The motive of this article is to predict the mood from the lyrics of a song. Word tokenization is used to make a list of words finding the importance as well as the frequency of the words from the lyrics. Natural Language Toolkit (NLTK) is a library written in python that has been used by the authors for word tokenization. NLTK has a module word-tokenize for word tokenization Notice that NLTK word tokenization also considers punctuation as a token. These unnecessary tokens are removed during the text-cleaning process of the dataset. Token count defines the length of the individual data. The researchers have set the maximum length 512.\par 

\begin{figure}[htpb]
\centering
\includegraphics[width=2.5in, height=1.5in]{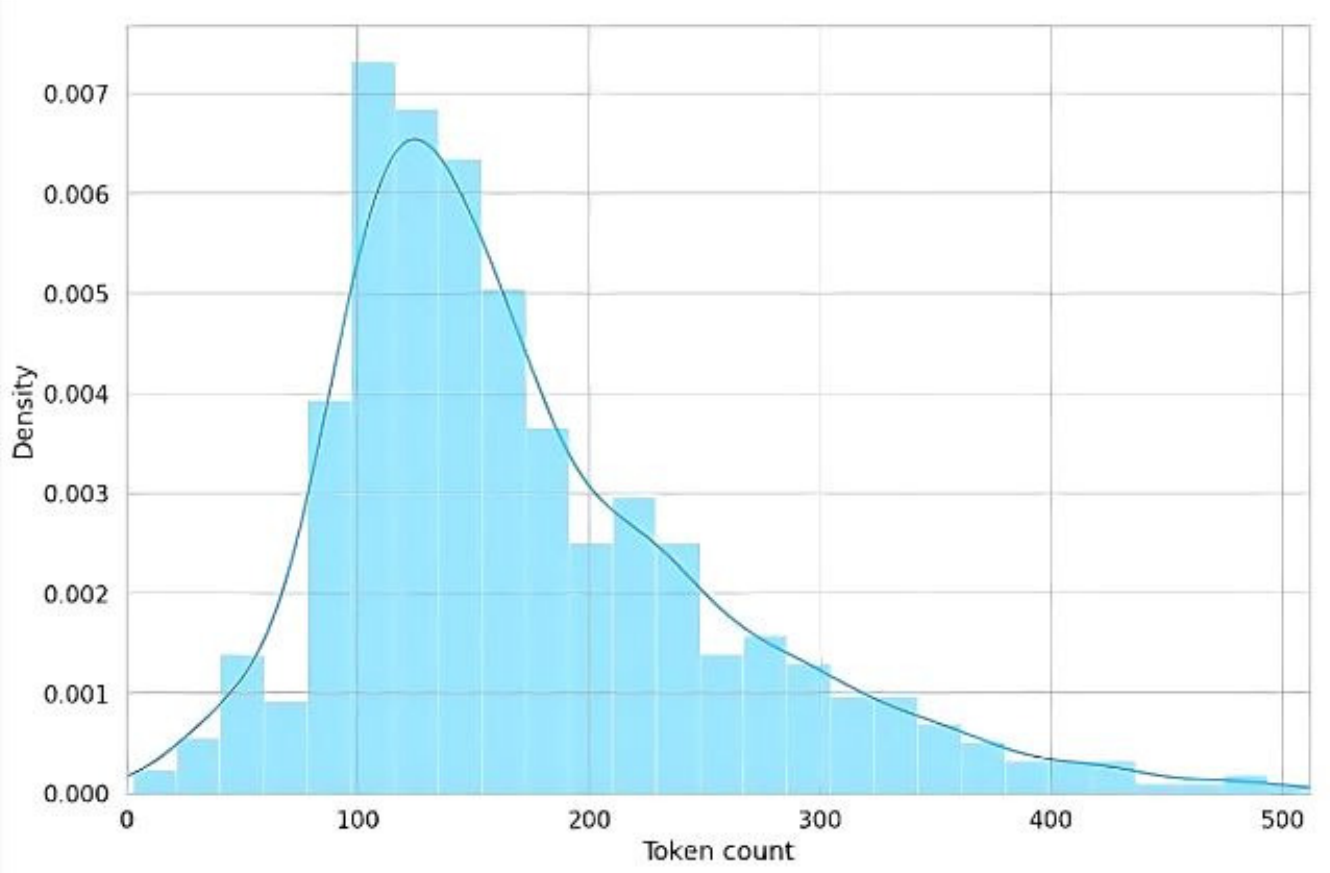}
\caption{Lexical Density in dataset}\label{fig:3}
\end{figure}

The authors have used NLTK's FreqDist class from the probability package to count tokens. After getting the number of tokens, the token-type ratio is determined. The authors also have determined the lexical density from the token count. Figure 3, it is shown how the density is reflected in token count. The highest density is gained when the unique token count is 100. The density is high when the token range is between 100-150. For other tokens, the density is going down. This result indicates there are few words to determine the mood and more common words.

\subsection{Deep Learning Approach}\label{subsec2}


BERT (Bidirectional Encoder representation from Transformers) is a new paper distributed by specialists at Google computer-based intelligence Language. It has made a ruckus to AI group of people introducing the best in its class brings about various
NLP(natural language processing) task, including question answering, Normal Language Surmising (MNLI), next sentence prediction, Musk language identification, and others \cite{bib14}.\par

The BERT key in \cite{bib15} is dedicated to development by applying the Bi-directional representation of the transformer, a mainstream considered model, to the visualization of language. This model has been created to expand several advantages over previous language models.  One of the advantages is the bidirectional approach BERT uses for
contextual understanding, which means that it reads the text both from left to right and from right to left. This results in a better understanding of the words and the context. BERT has the advantage of learning a wide range of language tasks as it is pre-trained on a large corpus of text. This pre-training also works well to understand the structure of a language. BERT works better with transfer learning- minimal data and specific NLP tasks, because of the pre-training on a large corpus. The model gives better performance on NLP tasks as it focuses on contextualized word embedding. Relying upon context made BERT achieve state-of-the-art performance on several NLP tasks, including natural language inference, sentiment analysis, and question-answering
answering. This research has chosen BERT as it is a powerful model including more flexibility to understand the words' context. \par
BERT uses a transformer, which is a consideration system that can learn the logical relationship between words or sub-words in learning content. In the transformer’s vanilla structure, the transformer contains two components — an encoder that examines the substance data and a decoder that conveys a gauge for the
task. Because BERT goes likely to create a model, simply the encoder system is fundamental. Google described Transformer’s feature-by-feature in \cite{bib16}.\par
Instead of a directional model that continuously reads content information (from left to right or from right to left), the Transformer encoder examines the entire gathering of words right away. This way is considered bidirectional, But it is more accurately said that is omnidirectional. This brand permits the model sample to get used to adjusting a word dependent on the entirety of its environmental factors like the left and right of the word. The schematic below is an undeniable level portrayal from
Transformer encoder. The information is a consecutive sequence of token counts, first inserted into a vector and then processed by the nervous system. The throughput is a series of H-size vectors, where each vector is compared to an information
token with similar files.
\subsection{Deep Learning Architecture}\label{subsec2}
In this research, BERT BASE Model has been used as the algorithm to train the dataset and predict songs' moods. To be more specific in order to work with the Bangla language a model called Bangla Bert Base is used. Along with the description of the BERT Model, a description of the NLP framework is given by the authors.\par

The researchers explained a transformer based model named BERT( Bidirectional Encoder Representations from Transformers) \cite{bib4}. The model is designed as a pre-trained deep bidirectional representations from the unlabeled text. The model works through jointing as well as conditioning on both the left and right contexts. As a result, the BERT model is fine-tuned with just one additional output layer to create state-of-the-art models for a wide range of NLP tasks.” BERT is a “deeply bidirectional” model. Bidirectional means learning information from both the left and the right side of a token’s context during the training phase. The bi-directionality of a model is important for
truly understanding the meaning of a language. \par
The BERT architecture builds on top of the Transformer in figure 4. BERT currently has two variants available: BERT Base, BERT Large. The mathematical model of BERT can be represented as a multi-layer bidirectional transformer encoder where each layer consists of two sub-layers: 
\begin{itemize}
  \item  A multi-head self-attention mechanism: It allows the model to attend to different parts of the input sequence to better capture dependencies between words.
  \item A position-wise feed-forward network: In each layer it applies a non-linear transformation to the output of the first layer to further refine the representations.
\end{itemize} 
BERT takes input of a sequence of tokenized words, and transform into embeddings through an embedding layer to pass through a series of encoder layers capturing contextual relationships between words. BERT also includes two special tokens: [CLS]-to the beginning of input sequence and [SEP] to seperate input sequence. 
Furthermore, a series of matrix multiplications
and non-linear transformations applied to the input sequence embeddings can represent the mathematical model of BERT used for NLP tasks.
\begin{figure}[htpb]
\centering
\includegraphics[width=2.5in, height=1.5in]{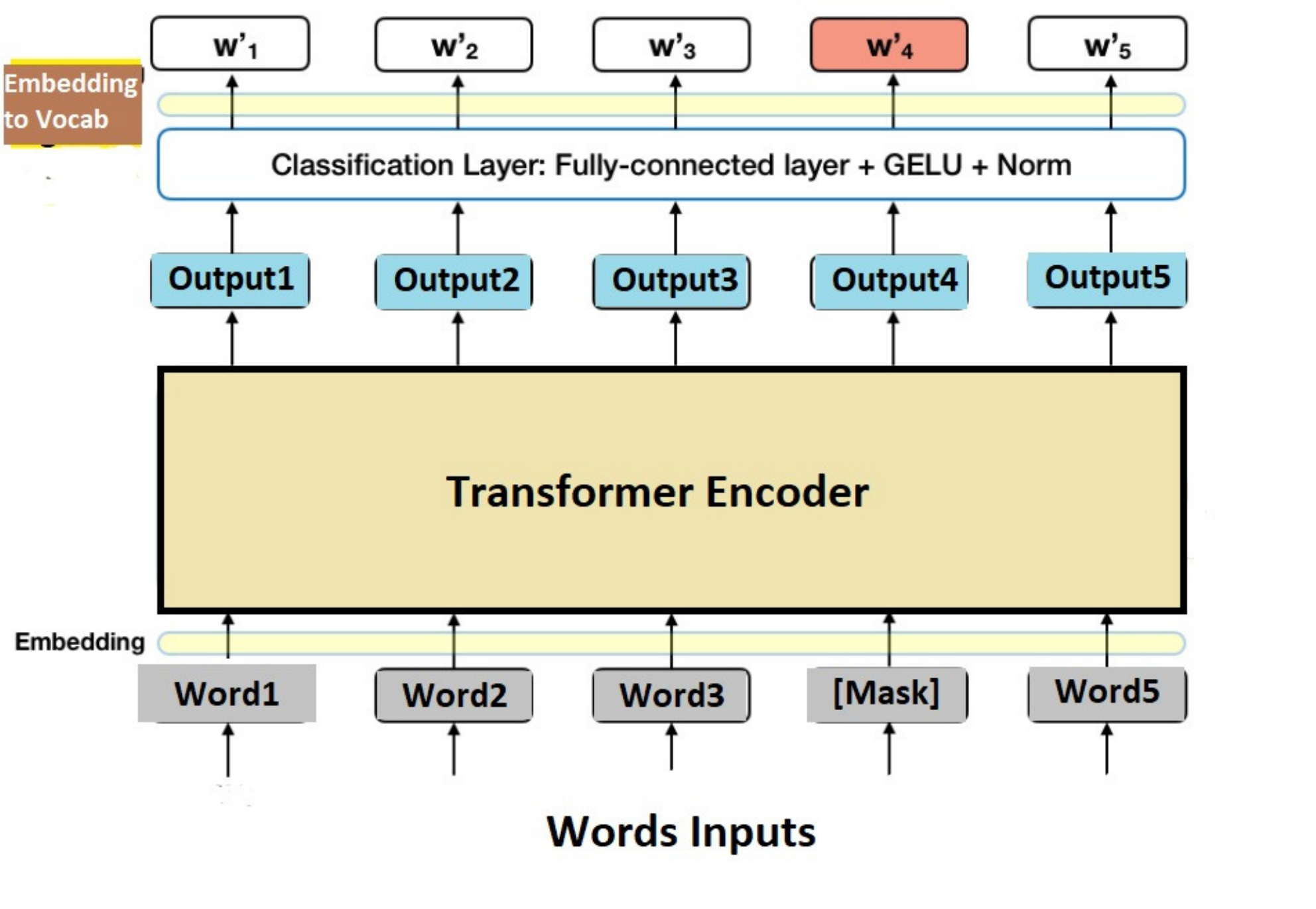}
\caption{Bert Architecture}\label{fig:5}
\end{figure}

For developing this research using a transformer library by hugging face. Transformers (some time ago known as PyTorch-transformers and PyTorch-pretrainedbert) give universally useful designs (BERT, GPT-2, Roberta, XLM, DistilBert, XLNet) for Natural Language Understanding (NLU) and Natural Language Generation (NLG) with over 32+ pre-trained models in over 100 dialects and profound interoperability between Jax, PyTorch, and TensorFlowFrom. This research has used BERT Base uncase in our model. This model contains 12 encoder layers, 768 feed-forward networks, and 12 attention heads. A special token [CLS] is provided for the first input token, the reason for which will become apparent later. CLS stands for classification here.\par

This research labeled all the song’s moods. Then in 2nd step, pre-processing of data which includes data cleaning, tokenizing, and padding has been done. After that, training of the data is done using Bangla BERT Base Model. After training data, evaluation, and testing of the data is performed.\par

\section{Result Analysis}\label{sec5}
This research is using Bert Base model to train the data and the prediction of the system is also based on the machine learning algorithm. For building this model a dataset of 4000 songs is prepared and labeled into four categories. The model is based on the moods of a song. A multi-class classifier is used by authors that can classify the lyrics' moods. The classifier appoints a large portion of the truly difficult work for BertModel.\par

A dropout class is used for certain regulations and an affiliated class is used only for the benefit. It is an important fact that the last layer has a rough yield because the unfortunate operation of cross-entropy in PyTorch is what it takes to work. This research applies a softmax function to the output to get the predicted probabilities of the trained model. In order to recreate the learning process from the BERT document, the authors have used Hugging Face’s AdamW optimizer. The researchers have corrected the weight loss so that it looks like the original document. Also in this article, a linear program is used without a heating step.
The BERT author made various suggestions with the purpose of fine-tuning \cite{bib17}:\par
\begin{table}[h]
\begin{center}
\begin{minipage}{174pt}
\caption{Suggestions with the purpose of fine-tuning}
\label{table1}
\begin{tabular}{|l|c|c|c|c|} 
   \hline
   \multicolumn{5}{|c|}{Fine-Tune} \\
   \hline
   Batch size & 8 & 16 & 32 & 64\\
   Learning rate (Adam) & 8e-5 & 5e-5 & 3e-5 & 2e-5\\
   Number of epochs & 40 & 100 & 200 & 300\\
   \hline
\end{tabular}
\end{minipage}
\end{center}
\end{table}

For training and validation, the authors have set the epochs to 100 with batch size 8. If the batch size is increased, it will significantly reduce the time of training the dataset, however, will reduce the training accuracy. There is a problem with google Collaboratory using the GPU. When the researcher increases the batch size, google collab gets out of memory. For this, this research uses a lesser batch size for training the model. Model training should be familiar with two exceptions: every time a package is loaded into the model, the scheduler is called. The researchers have used a clip-grad norm to cut the gradient of the model to avoid gradient explosion. At the training loop, the training history is sorted. After completing the training and validation process, storing the better model represented by the rate of the highest value of validation accuracy. \par

 In figure 5 the training accuracy and the validation accuracy of the dataset are described. In this research, accuracy has been gained during the training of the dataset. The figure represents the comparison between the training period and the validation period. After almost 7 epochs while training the dataset, the dataset's training accuracy starts to approach almost 100 percent. It gains the highest accuracy when the epoch is 12. On the other hand, the
validation accuracy is much lower than the training accuracy. It is quite unbalanced till 16 epochs. The validation accuracy becomes constant at 63 percent after that. The parameter has been fine-tuned to get an average accuracy of 65 percent.\par
\begin{figure}[htpb]
\centering
\includegraphics[width=3in, height=2.5in]{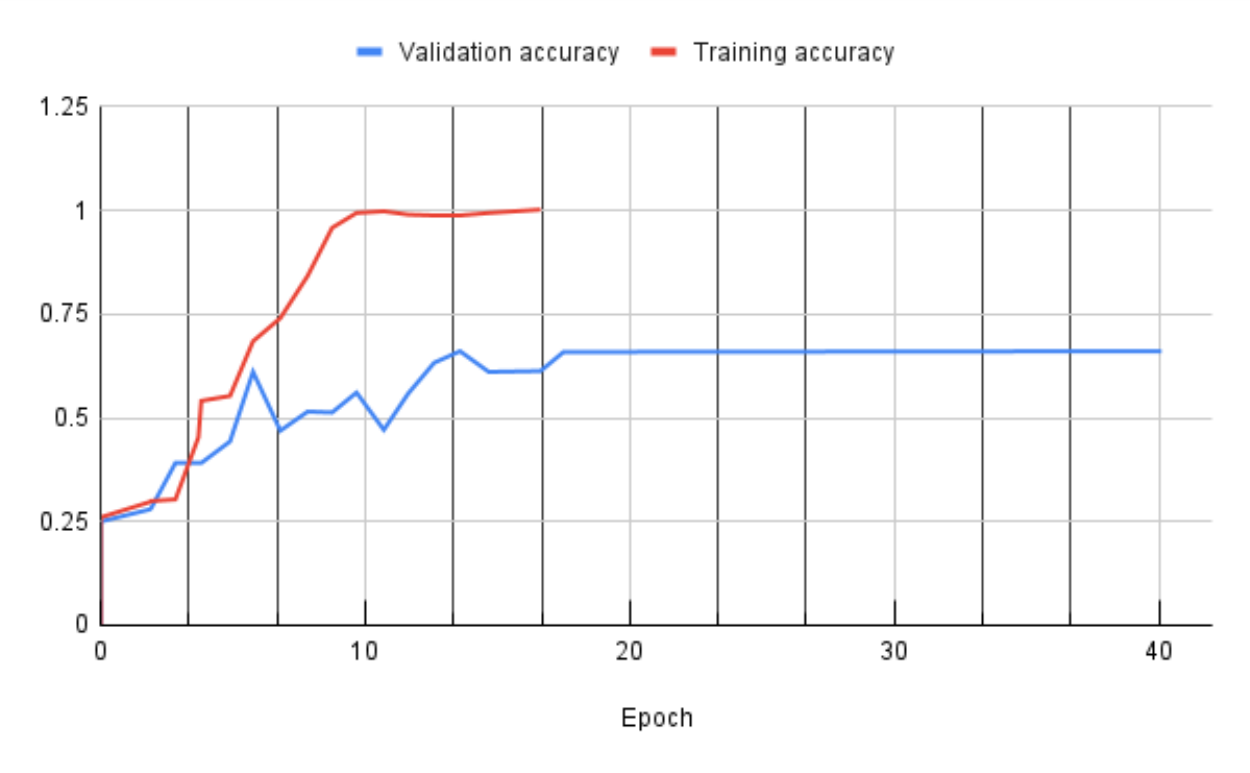}
\caption{Training and validation accuracy graph}\label{fig:5}
\end{figure}
However, this problem is occurring because of our data unbalance when training the model and for the batch size. When
increasing the batch size the google collab is getting out of memory. This is a common issue found out every moment during this research. When the batch size is decreased, everything is
working perfectly but it’s made in the model accuracy. The result shows an accuracy of 65 percent based on validation accuracy for the model.
\begin{figure}[htpb]
\centering
\includegraphics[width=2.5in, height=1.5in]{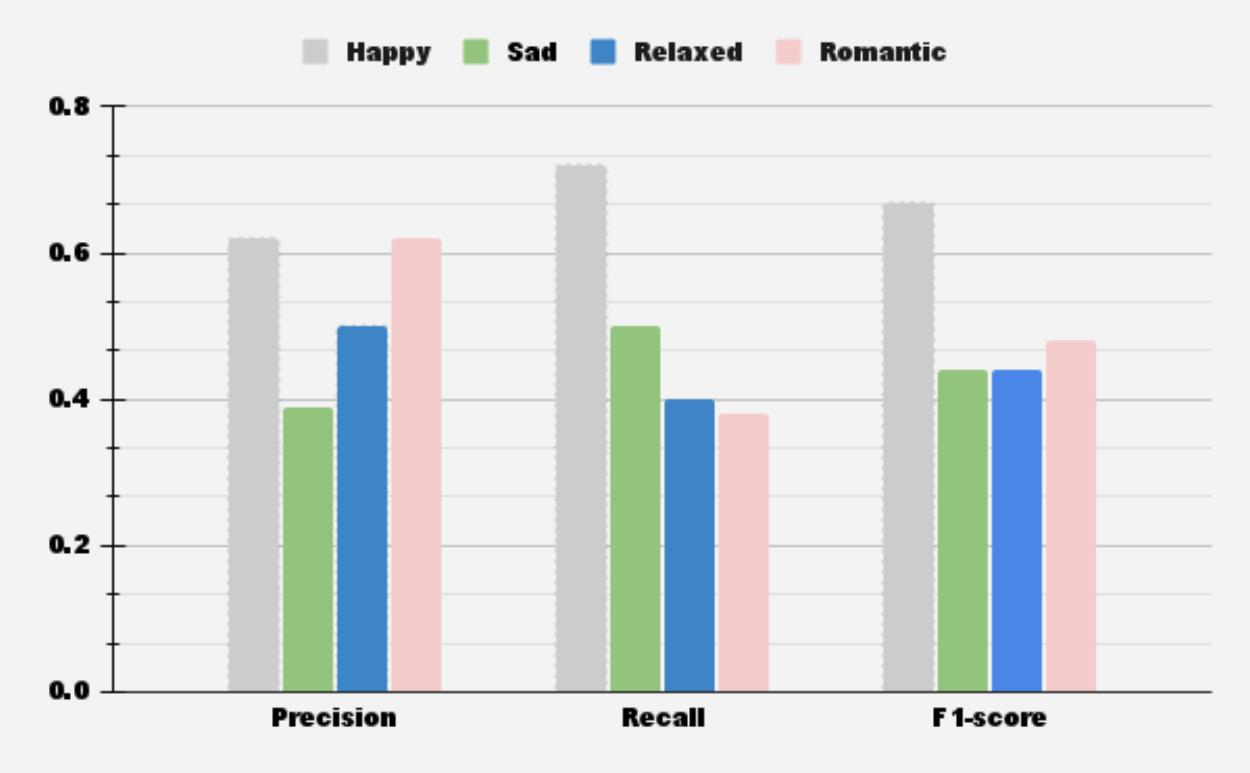}
\caption{Training and validation accuracy graph}\label{fig:6}
\end{figure}
The classification report of the model is portrayed in figure 6. The happy and romantic moods have the same amount of precision values. Though the recall value and f1-score of happy mood indicate the identification of this mood has more high accuracy than other moods. The identification of the relaxed mood has the minimum accuracy in the prediction system. From the graph, it’s quite difficult to identify the romantic, sad, and relaxed. This confusion will be cleared by the confusion matrix. In the confusion matrix class sad and romantic are at roughly equal frequency. Also sad and romantic class has an equal frequency to the happy class. By confusion matrix, the researchers confirm that the model has difficulty classifying the moods of romantic and relaxation songs.
\begin{figure}[htpb]
\centering
\includegraphics[width=3.3in, height=3in]{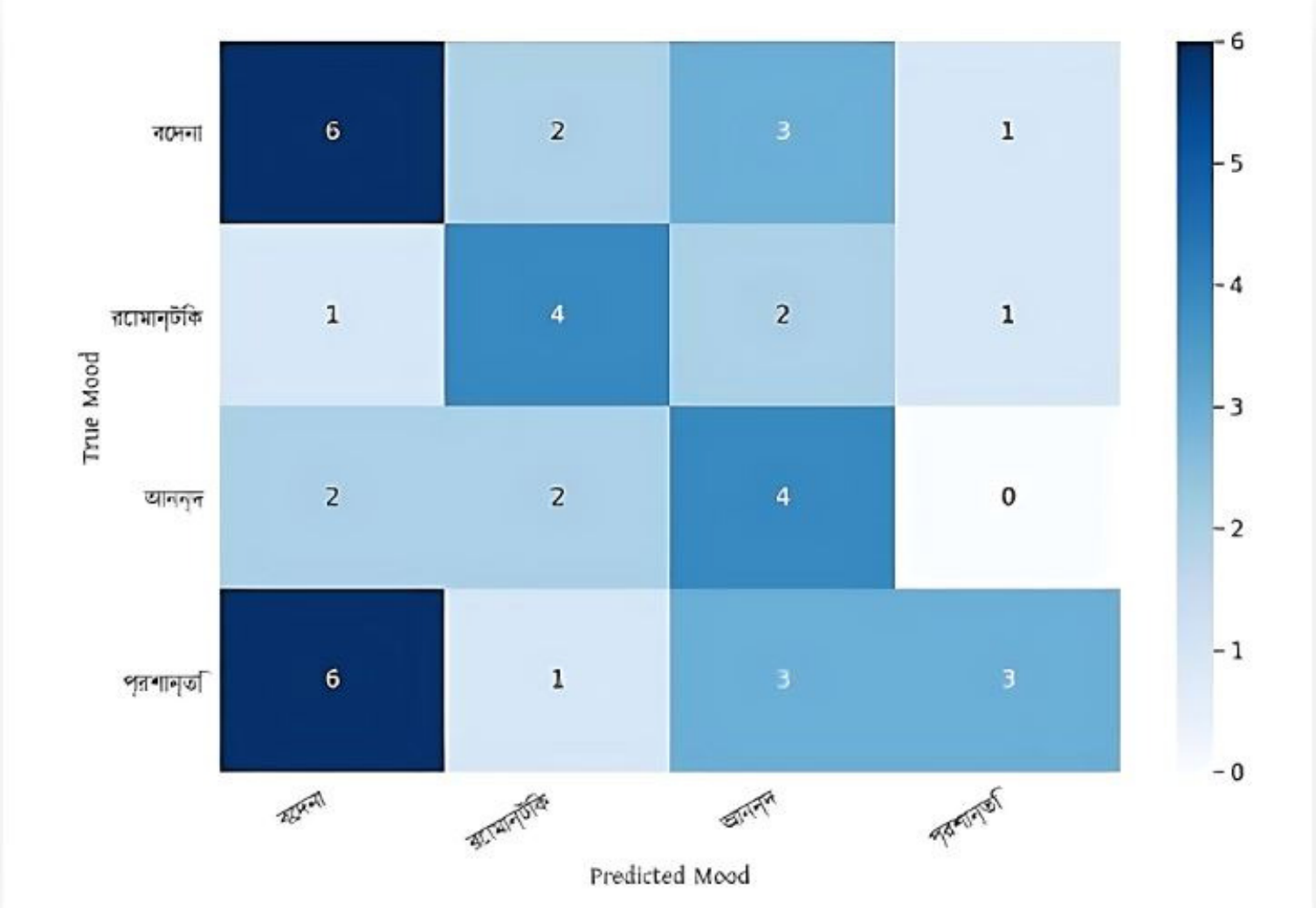}
\caption{Confusion Matrix}\label{fig:8}
\end{figure}
When the model is created this can identify the mood of a song’s lyrics. For Model Performance, the authors have trained the dataset and analyzed the model by validation of the dataset. This research creates the helper function for training and evaluating the model. 
\section{Conclusion}\label{6}
An extensive study has been carried out in this paper in relation to the use of machine learning tools in the field of Bangla Music Analysis. Though music analysis-related research is a very common research area for much linguistic research, the Bangla language hasn't been enriched in this case. Bangla songs are very popular both in Bangladesh and in many other countries. This research attempts to make a feasible system for listeners and people related to Bangla music. The collection of huge amounts of data and focus on the prediction of different moods as well as different emotions of Bangla music makes this research unique and useful. It has been shown that the moods can be identified in the songs using machine learning algorithms and justify the requirements of the users. This article will also show the change of Bangla music as well as the demand of people regarding music mood. In the future, the researchers would like to extend this model an make an application including more data making the system for informative and useful for people.
\bibliography{sn-bibliography}


\begin{thebibliography}{0}
\ifx \bisbn   \undefined \def \bisbn  #1{ISBN #1}\fi
\ifx \binits  \undefined \def \binits#1{#1}\fi
\ifx \bauthor  \undefined \def \bauthor#1{#1}\fi
\ifx \batitle  \undefined \def \batitle#1{#1}\fi
\ifx \bjtitle  \undefined \def \bjtitle#1{#1}\fi
\ifx \bvolume  \undefined \def \bvolume#1{\textbf{#1}}\fi
\ifx \byear  \undefined \def \byear#1{#1}\fi
\ifx \bissue  \undefined \def \bissue#1{#1}\fi
\ifx \bfpage  \undefined \def \bfpage#1{#1}\fi
\ifx \blpage  \undefined \def \blpage #1{#1}\fi
\ifx \burl  \undefined \def \burl#1{\textsf{#1}}\fi
\ifx \doiurl  \undefined \def \doiurl#1{\url{https://doi.org/#1}}\fi
\ifx \betal  \undefined \def \betal{\textit{et al.}}\fi
\ifx \binstitute  \undefined \def \binstitute#1{#1}\fi
\ifx \binstitutionaled  \undefined \def \binstitutionaled#1{#1}\fi
\ifx \bctitle  \undefined \def \bctitle#1{#1}\fi
\ifx \beditor  \undefined \def \beditor#1{#1}\fi
\ifx \bpublisher  \undefined \def \bpublisher#1{#1}\fi
\ifx \bbtitle  \undefined \def \bbtitle#1{#1}\fi
\ifx \bedition  \undefined \def \bedition#1{#1}\fi
\ifx \bseriesno  \undefined \def \bseriesno#1{#1}\fi
\ifx \blocation  \undefined \def \blocation#1{#1}\fi
\ifx \bsertitle  \undefined \def \bsertitle#1{#1}\fi
\ifx \bsnm \undefined \def \bsnm#1{#1}\fi
\ifx \bsuffix \undefined \def \bsuffix#1{#1}\fi
\ifx \bparticle \undefined \def \bparticle#1{#1}\fi
\ifx \barticle \undefined \def \barticle#1{#1}\fi
\bibcommenthead
\ifx \bconfdate \undefined \def \bconfdate #1{#1}\fi
\ifx \botherref \undefined \def \botherref #1{#1}\fi
\ifx \url \undefined \def \url#1{\textsf{#1}}\fi
\ifx \bchapter \undefined \def \bchapter#1{#1}\fi
\ifx \bbook \undefined \def \bbook#1{#1}\fi
\ifx \bcomment \undefined \def \bcomment#1{#1}\fi
\ifx \oauthor \undefined \def \oauthor#1{#1}\fi
\ifx \citeauthoryear \undefined \def \citeauthoryear#1{#1}\fi
\ifx \endbibitem  \undefined \def \endbibitem {}\fi
\ifx \bconflocation  \undefined \def \bconflocation#1{#1}\fi
\ifx \arxivurl  \undefined \def \arxivurl#1{\textsf{#1}}\fi
\csname PreBibitemsHook\endcsname

\end{thebibliography}


\begin{thebibliography}{17}
\bibitem{bib1}Zurawicki, L. Neuromarketing: Exploring the brain of the consumer.  (2010)
\bibitem{bib2}Schedl, M., Zamani, H., Chen, C., Deldjoo, Y. \& Elahi, M. Current challenges and visions in music recommender systems research. {\em International Journal Of Multimedia Information Retrieval}. \textbf{7} pp. 95-116 (2018)
\bibitem{bib3}Nummenmaa, L., Putkinen, V. \& Sams, M. Social pleasures of music. {\em Current Opinion In Behavioral Sciences}. \textbf{39} pp. 196-202 (2021)
\bibitem{bib4}Wang, Z. \& Xia, G. MuseBERT: Pre-training Music Representation for Music Understanding and Controllable Generation. {\em 22nd International Society For Music Information Retrieval Conference, ISMIR 2021, Online, November 7-12, 2021}. pp. 722-729 (2021)
\bibitem{bib5}Walaa Medhat, H. Sentiment analysis algorithms and applications: A survey. {\em  Ain Shams Engineering Journal}. \textbf{5}, 1093-1113 (2014)
\bibitem{bib6}Doaa Mohey El-Din Mohamed Hussein A survey on sentiment analysis challenges,. {\em Journal Of King Saud University - Engineering Sciences,}. \textbf{30} pp. 330-338 (2018)
\bibitem{bib7}Rabeya, T., Chakraborty, N., Ferdous, S., Dash, M. \& Al Marouf, A. Sentiment Analysis of Bangla Song Review- A Lexicon Based Backtracking Approach. {\em 2019 IEEE International Conference On Electrical, Computer And Communication Technologies (ICECCT)}. pp. 1-7 (2019)
\bibitem{bib8}Mamun, M., Kadir, I., RABBY, A. \& Azmi, A. Bangla Music Genre Classification Using Neural Network. {\em 2019 8th International Conference System Modeling And Advancement In Research Trends (SMART)}. pp. 397-403 (2019)
\bibitem{bib9}Zaanen, M. \& Kanters, P. Automatic Mood Classification Using TF*IDF Based on Lyrics.. {\em Proceedings Of The 11th International Society For Music Information Retrieval Conference, ISMIR 2010}. pp. 75-80 (2010,1)
\bibitem{bib10}Kashyap, N., Choudhury, T., Chaudhary, D. \& Lal, R. Mood Based Classification of Music by Analyzing Lyrical Data Using Text Mining.  (2016,9)
\bibitem{bib11}Raschka, S. MusicMood: Predicting the mood of music from song lyrics using machine learning. {\em ArXiv}. \textbf{abs/1611.00138} (2016)
\bibitem{bib12}Urmi, N., Ahmed, N., Sifat, M., Islam, S. \& Jameel, A. BanglaMusicMooD: A Music Mood Classifier from Bangla Music Lyrics.  (2021,1)
\bibitem{bib13}Çano, E. \& Morisio, M. Music Mood Dataset Creation Based on Last.fm Tags.  (2017)
\bibitem{bib14}Devlin, J., Chang, M., Lee, K. \& Toutanova, K. BERT: Pre-training of Deep Bidirectional Transformers for Language Understanding. {\em ArXiv}. \textbf{abs/1810.04805} (2019)
\bibitem{bib15}Kaliyar, R. A Multi-layer Bidirectional Transformer Encoder for Pre-trained Word Embedding: A Survey of BERT.  2020 10th International Conference On Cloud Computing, Data Science \& Engineering. pp. 336-340 (2020)
\bibitem{bib16}Wolf, T., Debut, L., Sanh, V., Chaumond, J., Delangue, C., Moi, A., Cistac, P., Rault, T., Louf, R., Funtowicz, M. \& Others Huggingface's transformers: State-of-the-art natural language processing. {\em ArXiv Preprint ArXiv:1910.03771}. (2019)
\bibitem{bib17}Chi, S., Qiu, X., Xu, Y. \& Huang, X. How to Fine-Tune BERT for Text Classification?.  (2019,5)
\end{thebibliography}

%
\end{document}